\documentclass[aps,prd,twocolumn,nofootinbib,superscriptaddress,preprintnumbers]{revtex4-2}
\usepackage{amsfonts,amsmath,amssymb,amsthm,braket,nicefrac}
\usepackage{booktabs,array}
\usepackage[utf8]{inputenc}
\usepackage{float,graphicx,hhline}
\usepackage{appendix}
\usepackage[dvipsnames]{xcolor}
\usepackage{mathtools}
\usepackage[caption=false]{subfig}
\usepackage[export]{adjustbox}
\usepackage[normalem]{ulem}

\AtBeginDocument{
\heavyrulewidth=.08em
\lightrulewidth=.05em
\cmidrulewidth=.03em
\belowrulesep=.65ex
\belowbottomsep=0pt
\aboverulesep=.4ex
\abovetopsep=0pt
\cmidrulesep=\doublerulesep
\cmidrulekern=.5em
\defaultaddspace=.5em
}

\usepackage{hyperref}
\hypersetup{
    pdftitle={Flow-based density of states for complex actions},
    colorlinks=true,     
    linkcolor=blue,      
    citecolor=blue,      
    filecolor=blue,      
    urlcolor=blue        
}
\usepackage[nameinlink]{cleveref}

\newcommand{\im}{\mathrm{i}}

\begin{document}

\title{Flow-based density of states for complex actions}

\author{Jan~M.~Pawlowski}
\affiliation{Institut f\"ur Theoretische Physik, Universit\"at Heidelberg, Philosophenweg 16, 69120 Heidelberg, Germany}
\author{Julian~M.~Urban}
\affiliation{Center for Theoretical Physics, Massachusetts Institute of Technology, Cambridge, MA 02139, USA}
\affiliation{The NSF AI Institute for Artificial Intelligence and Fundamental Interactions}

\preprint{MIT-CTP/5517}

\begin{abstract}
    Emerging sampling algorithms based on normalizing flows have the potential to solve ergodicity problems in lattice calculations. Furthermore, it has been noted that flows can be used to compute thermodynamic quantities which are difficult to access with traditional methods. This suggests that they are also applicable to the density-of-states approach to complex action problems. In particular, flow-based sampling may be used to compute the density directly, in contradistinction to the conventional strategy of reconstructing it via measuring and integrating the derivative of its logarithm. By circumventing this procedure, the accumulation of errors from the numerical integration is avoided completely and the overall normalization factor can be determined explicitly. In this proof-of-principle study, we demonstrate our method in the context of two-component scalar field theory where the internal $O(2)$ symmetry is explicitly broken by an imaginary external field. First, we concentrate on the zero-dimensional case which can be solved exactly. We show that with our method, the Lee-Yang zeroes of the associated partition function can be successfully located. Subsequently, we confirm that the flow-based approach correctly reproduces the density computed with conventional methods in one- and two-dimensional models.
\end{abstract}
\maketitle

\date{\today}


\section{Introduction}
\label{sec:intro}

Lattice calculations are a powerful approach to study quantum field theories non-perturbatively by applying Markov Chain Monte Carlo (MCMC) sampling~\cite{Morningstar:2007zm}, see~\cite{Brower:2019oor, Lehner:2019wvv, Kronfeld:2019nfb, Cirigliano:2019jig, Detmold:2019ghl, Bazavov:2019lgz, Mathur:2016cko, Joo:2019byq} for recent reviews. However, for many physically interesting cases, the associated Euclidean lattice action is complex-valued. This prohibits the application of standard importance sampling, the most prominent example being the notorious sign problem in lattice QCD at finite chemical potential~\cite{Aarts:2015tyj, Gattringer:2016kco}. In this context, it has been shown that with the density-of-states (DoS) approach~\cite{Bhanot:1986hi, Bhanot:1986kv, Bhanot:1986ku, Bhanot:1987bu, Bhanot:1987nv, Wang:2000fzi, Fodor:2007vv, Langfeld:2012ah, Langfeld:2013xbf, Langfeld:2015fua, Borsanyi:2021gqg}, certain complex action problems can be successfully treated~\cite{Langfeld:2014nta, Gattringer:2015lra, Gattringer:2015eey, Langfeld:2016kty, Garron:2016noc, Korner:2020vjw, Lucini:2021nft}. However, directly computing the DoS is generally not possible due to the intrinsically high variance of the associated observables. Instead, the usual approach is to measure the derivative of its logarithm with restricted MCMC calculations, followed by reconstructing the DoS itself via numerical integration. The high precision required to control the accumulation of errors from the approximation of the integral can be computationally expensive.

Recently, it has been noted that similar thermodynamic quantities in lattice field theory can be computed directly using generative machine learning models with tractable probability densities~\cite{Nicoli:2019gun, Nicoli:2020njz, Nicoli:2021inv}, thereby completely avoiding the aforementioned numerical reconstruction of the quantity of interest. Normalizing flows are one such class of probabilistic models for which both efficient sampling and density estimation are made possible using a change-of-variables formula~\cite{tabak2010, tabak2013, rezende2016variational, dinh2017density, JMLR:v22:19-1028}. They have been successfully applied to model real scalar field theory~\cite{Albergo:2019eim, Hackett:2021idh, DelDebbio:2021qwf, Caselle:2022acb, Matthews:2022sds}, $U(1)$ and $SU(N)$ gauge theories~\cite{Kanwar:2020xzo, Boyda:2020hsi, Foreman:2021ixr, Foreman:2021rhs, Foreman:2021ljl}, as well as theories with dynamical fermions~\cite{Albergo:2021bna, Finkenrath:2022ogg, Albergo:2022qfi}. For their application to sign problems, flows have been studied in the context of contour deformations~\cite{Lawrence:2021izu, Rodekamp:2022xpf}.

In this work, we apply flow-based sampling to the direct computation of the DoS for lattice field theories with complex actions. Specifically, we investigate scalar $\phi^4$-theory with two real-valued components where one is coupled to an imaginary external field, thereby explicitly breaking the internal $O(2)$ symmetry. We first consider the exactly solvable, zero-dimensional case as a toy model for a proof-of-principle demonstration, showing that the DoS as well as the partition function and magnetization as functions of the external field are computed correctly with our approach. In particular, we can locate Lee-Yang zeroes~\cite{Lee:1952ig} of the partition function together with the associated discontinuities in the magnetization. We then apply the approach to actual lattice models in one and two dimensions, accurately reproducing the densities obtained with conventional MCMC methods.

This paper is organized as follows. In \Cref{sec:dos}, we briefly review the DoS approach pertinent to the type of complex action problem considered here. \Cref{sec:flows} serves to introduce the basic concepts of normalizing flows that are relevant to this work. We explain our approach in \Cref{sec:flow-dos} and present numerical results in \Cref{sec:results}. We summarize our contributions and provide an outlook in \Cref{sec:summary}.


\section{Density of states}
\label{sec:dos}

We consider lattice field theories with complex-valued actions of the form
\begin{equation}
    S(\phi) = S_r(\phi) + \im h X(\phi) \ ,
\end{equation}
where $S_r,X,h \in \mathbb{R}$. The partition function and expectation values of observables are defined as
\begin{equation}\label{eq:Z}
    Z = \int \mathcal{D}\phi\, e^{-S_r(\phi) - \im h X(\phi)} \ ,
\end{equation}
\begin{equation}
    \langle \mathcal{O} \rangle = \frac{1}{Z} \int \mathcal{D}\phi\, e^{-S_r(\phi) - \im h X(\phi)} \mathcal{O}(\phi) \ .
\end{equation}
Since the action is complex, standard importance sampling is not directly applicable and reweighting often becomes prohibitively expensive when increasing $h$ due to the average phase factor being close to zero.

One ansatz to make the computation more tractable is to consider the DoS as a function of the quantity that generates the imaginary part of the action, i.e.
\begin{equation}\label{eq:rho}
    \rho(c) = \int \mathcal{D}\phi\, e^{-S_r(\phi)} \delta(X(\phi) - c) \ .
\end{equation}
Essentially, $\rho(c)$ corresponds to slices of the partition function for the real part of the action, with the configuration space restricted to hypersurfaces of constant $X(\phi) = c$. In MCMC calculations, this restriction can be achieved e.g.\ by confining the dynamics through additional rejections, or by replacing the $\delta$-distribution with a Gaussian of finite width, which is the approach used in the present work (see \Cref{sec:flow-dos} for details).

If $\rho(c)$ is known, the partition function for the full action as well as expectation values of observables (that are functions of $c$ only) can be computed in terms of one-dimensional integrals with a residual phase,
\begin{equation}\label{eq:rho-Z}
    Z = \int \mathrm{d}c\, \rho(c)\, e^{-\im h c} \ ,
\end{equation}
\begin{equation}\label{eq:rho-obs}
    \langle \mathcal{O} \rangle = \frac{1}{Z} \int \mathrm{d}c\, \rho(c)\, e^{-\im h c}\, \mathcal{O}(c) \ .
\end{equation}
However, similar to partition functions themselves and thermodynamic quantities in general, a direct computation of $\rho(c)$ is often infeasible with conventional MCMC algorithms due to the high variance associated with the required observables. Instead, it is usually reconstructed from measurements of $\partial_c \log \rho(c)$, as detailed below.


\section{Normalizing flows}
\label{sec:flows}

Starting with a prior distribution over a continuous space $\mathcal{X}$ with a known probability density $r(\xi)$, an invertible transport map (``flow'') $f: \mathcal{X} \to \mathcal{X}, \xi \mapsto \phi$ can be used to redistribute samples under $r$ to samples under a new density $q(\phi)$. We only require that the map be diffeomorphic, i.e.~that both $f$ and its inverse are differentiable. The resulting density $q(\phi)$ is fixed by the choice of prior distribution and map, and it can be evaluated explicitly as
\begin{equation}
    q(\phi) = r(\xi) \left|\det\left(\frac{\partial f}{\partial \xi}\right)\right|^{-1} \ ,
\end{equation}
where $\phi = f(\xi)$ and $\det(\partial f / \partial \xi)$ is the Jacobian determinant of $f$. Because the density after the transformation can be computed explicitly, flows provide a mechanism for both sampling and density estimation.

By choosing a sufficiently expressive parametrization of $f$, the space of associated transformations---corresponding to a large variational family of model densities $q$---can be explored through numerical optimization in order to find an instance that best approximates some target density $p$. In particular, the parameters of $f$ may be optimized by performing stochastic gradient descent on a measure of the discrepancy between the two densities $q$ and $p$, i.e.\ an appropriate loss function.

\begin{figure}
    \centering
    \includegraphics[width=0.8\linewidth]{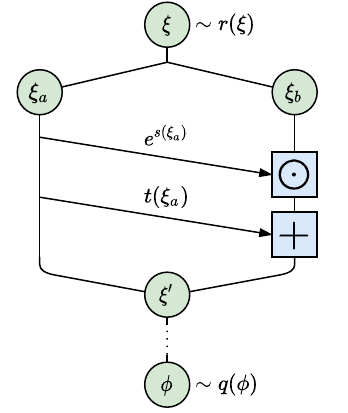}
    \caption{Illustration of the affine coupling layer defined in \Cref{eq:affine}. The blue boxes depict element-wise operations.}
    \label{fig:affine}
\end{figure}

A common choice is the Kullback-Leibler (KL) divergence, which is a measure of the relative entropy between distributions. It is defined as
\begin{equation}
\begin{aligned}
    D_{\mathrm{KL}}(q||p) &= \int \mathcal{D}\phi\, q(\phi) (\log q(\phi) - \log p(\phi)) \\
    &= \big\langle \log q(\phi) - \log p(\phi) \big\rangle_{\phi \sim q(\phi)} \geq 0 \ ,
\end{aligned}
\end{equation}
and takes the minimum value $D_{\mathrm{KL}}(q||p) = 0$ iff $q = p$. With expectation values measured using samples from the model distribution $q$, $D_{\mathrm{KL}}(q||p)$ can be stochastically estimated without requiring samples from the target distribution $p$.

For targets of the form $p(\phi) = e^{-S(\phi)} / Z$, the KL divergence takes the form
\begin{equation}
    D_{\mathrm{KL}}(q||p) = \big\langle \log q(\phi) + S(\phi) \big\rangle_{\phi \sim q(\phi)} + \log Z \ .
\end{equation}
Since the partition function $Z$ is usually not known a priori, $D_{\mathrm{KL}}$ can only be estimated up to the constant $\log Z$. However, this does not affect gradients and one may freely use $(D_{\mathrm{KL}} - \log Z)$ as the loss function for optimization, which then provides a bound on $\log Z$. Writing the partition function as
\begin{equation}\label{eq:flow-Z}
    Z = \int \mathcal{D}\phi \, q(\phi) \frac{e^{-S(\phi)}}{q(\phi)} = \left\langle e^{-S(\phi) - \log q(\phi)} \right\rangle_{\phi \sim q(\phi)} \ ,
\end{equation}
it follows that any model giving good agreement to the target distribution necessarily provides a precise, unbiased estimate of $Z$ at low cost through model samples alone, without the need to ever sample from the target.

A common building block for the construction of invertible flow transformations is the affine coupling layer. In each such layer, the input $\xi$ is split into two equal-sized subsets $\xi_a, \xi_b$ which are transformed according to
\begin{equation}\label{eq:affine}
\begin{aligned}
    \xi_a' &= \xi_a \\
    \xi_b' &= \xi_b \odot e^{s(\xi_a)} + t(\xi_a) \ ,
\end{aligned}
\end{equation}
i.e.\ $\xi_a$ remains unchanged (``frozen'') while $\xi_b$ is updated (``active''); see \Cref{fig:affine} for an illustration. Here, the symbol $\odot$ denotes element-wise multiplication. Each affine coupling layer is trivially invertible and has a triangular Jacobian matrix, thereby making the computation of its determinant and thus the model density $q$ tractable. The context functions $s,t$ that take the frozen variables as inputs and are used to update the active ones can be arbitrary functions. They are commonly parametrized by deep, feedforward neural networks, with are then trained during optimization. Importantly, they are not required to be invertible themselves, thereby providing much freedom in choosing a particular parametrization. Expressive flow transformations are built by chaining together many such affine coupling layers with alternating frozen and active subsets.


\section{Flow-based density of states}
\label{sec:flow-dos}

As already mentioned in \Cref{sec:dos}, we employ a formulation of the DoS approach where the $\delta$-distribution in \Cref{eq:rho} is replaced by a Gaussian of finite width, following e.g.~\cite{Fodor:2007vv, Borsanyi:2021gqg}. This enables the straightforward application of both standard sampling algorithms like Hybrid/Hamiltonian Monte Carlo (HMC) as well as our flow-based approach. Exactness of all expressions can be retained at the cost of a residual sign problem (which is tractable for sufficiently small width) or by extrapolating to the limit of vanishing width.

First, we note that the result of the Gaussian integral
\begin{equation}
    \int \mathrm{d}c \, e^{-\frac{P}{2}(c - a)^2} = \sqrt{\frac{2\pi}{P}} \equiv \mathcal{N}
\end{equation}
is independent of $a$. Hence, we can rewrite \Cref{eq:Z} as
\begin{equation}
    Z = \int \mathcal{D}\phi \int \mathrm{d}c \, e^{-\frac{P}{2}(c - X(\phi))^2 - \log \mathcal{N} - S_r(\phi) - \im h X(\phi)} \ .
\end{equation}
We then define the $P$-dependent DoS as
\begin{equation}\label{eq:rho-P}
    \rho_P(c) = \int \mathcal{D}\phi \, e^{-S_{c,P}(\phi)} \ ,
\end{equation}
where
\begin{equation}
    S_{c,P}(\phi) = S_r(\phi) + \frac{P}{2} (c - X(\phi))^2 + \log \mathcal{N} \ .
\end{equation}
The ``true'' DoS as defined in \Cref{eq:rho} is recovered in the limit $P \longrightarrow \infty$.

Assuming continuity and convergence of the integrals, the partition function can be expressed in terms of $\rho_P$ as
\begin{equation}\label{eq:rho-P-Z}
\begin{aligned}
    Z &= \int \mathrm{d}c \int \mathcal{D}\phi \, e^{-S_{c,P}(\phi) - \im h X(\phi)} \\
    &= \int \mathrm{d}c \, \rho_P(c) \frac{\int \mathcal{D}\phi \, e^{-S_{c,P}(\phi) - \im h X(\phi)}}{\int \mathcal{D}\phi \, e^{-S_{c,P}(\phi)}} \\
    &= \int \mathrm{d}c \, \rho_P(c) \big\langle e^{-\im h X(\phi)} \big\rangle_{\phi \sim e^{-S_{c,P}(\phi)}} \ ,
\end{aligned}
\end{equation}
Hence, in this formulation, the partition function is still a one-dimensional integral over the $P$-dependent DoS, but with an additional average phase factor computed on ensembles sampled with $S_{c,P}(\phi)$. The fluctuations of this phase factor are tractable as long as the parameter $P$ is large enough, such that $X(\phi)$ does not deviate too strongly from $c$. Accordingly, expectation values of observables can be written as
\begin{equation}\label{eq:rho-P-obs}
    \langle \mathcal{O} \rangle = \frac{\int \mathrm{d}c \, \rho_P(c) \big\langle e^{-\im h X(\phi)} \mathcal{O}(\phi) \big\rangle_{\phi \sim e^{-S_{c,P}(\phi)}}}{\int \mathrm{d}c \, \rho_P(c) \big\langle e^{-\im h X(\phi)} \big\rangle_{\phi \sim e^{-S_{c,P}(\phi)}}} \ .
\end{equation}

As mentioned previously, a direct computation of $\rho_P(c)$ with traditional MCMC methods is often infeasible for problems of interest. Instead, the usual strategy is to compute
\begin{equation}
\begin{aligned}
    \frac{\partial \log \rho_P(c)}{\partial c} &= \frac{\int \mathcal{D}\phi \, e^{-S_{c,P}(\phi)} (-P(c - X(\phi)))}{\int \mathcal{D}\phi \, e^{-S_{c,P}(\phi)}} \\
    &= \big\langle -P(c - X(\phi)) \big\rangle_{\phi \sim e^{-S_{c,P}(\phi)}} \ ,
\end{aligned}
\end{equation}
and then to reconstruct $\log(\rho_P(c) / \rho_P(0))$ by numerical integration, e.g.\ with the trapezoidal rule. In contrast, normalizing flows trained with $S_{c,P}(\phi)$ as the target action allow for a direct computation of $\rho_P(c)$ (including the overall factor $\rho_P(0)$) using configurations sampled from $q(\phi)$, as long as the overlap of the target and model distributions is sufficient. This can be seen by rewriting \Cref{eq:rho-P} as
\begin{equation}
\begin{aligned}
    \rho_P(c) &= \int \mathcal{D}\phi \, q(\phi) \frac{e^{-S_{c,P}(\phi)}}{q(\phi)} \\
    &= \left\langle e^{-S_{c,P}(\phi) - \log q(\phi)} \right\rangle_{\phi \sim q(\phi)} \ ,
\end{aligned}
\end{equation}
similar to \Cref{eq:flow-Z}. A successfully trained flow minimizes the fluctuations of the exponent in the last expression, such that the variance of the expectation value remains tractable. This is precisely the crucial advantage of flow-based sampling over conventional MCMC methods that allows the computation of thermodynamic quantities via variationally optimized reweighting~\cite{Nicoli:2020njz, Nicoli:2021inv}.

In order to compute $\rho_P$ across a wide range, one could train independent flows for each value of $c$. Alternatively, a more efficient approach would be to start by training one flow at some given point (e.g.\ $c = 0$) and then perform retraining for each additional point. However, since high precision in $c$ is desired, these strategies seem impractical. Apart from such a training procedure already being computationally expensive, a large number of different parameter sets for all the individual flow transformations would then have to be stored and loaded into memory for evaluation. Instead, we propose to encode the full information about $\rho_P$ for all $c$ in a single flow model. This is achieved by promoting the transport map $f(\xi)$ to a conditional transformation $f_c(\xi)$, which additionally depends on $c$. In particular, the context functions $s,t$ of all affine couplings as defined in \Cref{eq:affine} are modified to take $c$ as an additional input. This only marginally increases the computational effort of evaluating the transformation, although it may be necessary to make the flow more expressive overall in order to properly model the dependence on $c$.

Furthermore, we introduce an additional $c$-dependent offset at the last layer, such that the conditional generation of field configurations $\phi$ from prior samples $\xi$ takes the form
\begin{equation}\label{eq:c-flow}
    \phi(\xi|c) = f_c(\xi) + \bar{\phi}(c) \ ,
\end{equation}
with $\bar{\phi}(c)$ chosen such that $X(\bar{\phi}(c)) = c$. This offset already provides the correct mean field configuration for each $c$ and thereby greatly simplifies training from the start, because the flow only has to model the distribution around the given $\bar{\phi}(c)$. Since this amounts to just a constant shift, the Jacobian of the transformation remains unchanged.

It should be stressed that for the simple architecture utilized in the present work, outputs of intermediate layers are not necessarily meaningful since we do not specify a particular flow trajectory in the space of probability distributions. Only the field variables at the prior and target endpoints of the transformation carry definite meaning. Nevertheless, enforcing such a particular sequence of intermediate densities by appropriate loss functions---e.g., by interpolating the coupling constants of non-quadratic action terms as well as the parameter $P$ from 0 to their target values, thereby resulting in a Gaussian prior on one side and the desired target density on the other side---could have several advantages, and will be considered in future work. Most importantly, such an approach makes the relation of the present approach to thermodynamic integration and the Jarzynski equality~\cite{PhysRevLett.78.2690} more concrete. This, in turn, enables the application of stochastic normalizing flow approaches~\cite{wu2020stochastic, Caselle:2022acb, Matthews:2022sds}, which may be particularly effective at determining the types of thermodynamic quantities considered here. Furthermore, explicitly determining the precise $P$-dependence of the DoS in this manner may allow a direct extrapolation to the vanishing-width limit without the need to optimize several flow instances for different target parameters.

Taken together, the full transformation defined in \Cref{eq:c-flow} consisting of a conditional transport map and an additional offset induces a conditional model distribution $q_c$ for each $c$, such that the $P$-dependent DoS may be computed as
\begin{equation}
    \rho_P(c) = \big\langle e^{-S_{c,P}(\phi) - \log q_c(\phi)} \big\rangle_{\phi \sim q_c(\phi)} \ .
\end{equation}
During training, one may estimate the conditional KL divergence $D_{\mathrm{KL}}(q_c||e^{-S_{c,P}})$ at random points $c$, distributed uniformly across a sufficiently large interval, in order to enforce optimal generalization for arbitrary $c$.

At this point we emphasize again that in order to evaluate the above expression for $\rho_P$, only samples from the model distribution are required. Importantly, this implies that once the flow has been trained, the remaining computations can be performed extremely efficiently in a manner that has been fittingly described as ``embarrassingly parallel''. In particular, field configurations do not need to be arranged in a Markov chain and no accept/reject steps are necessary. This constitutes a further potential advantage of our proposed approach over conventional MCMC calculations.


\section{Numerical Results}
\label{sec:results}

\subsection{Zero-Dimensional Model}

\begin{figure}
    \centering
    \includegraphics{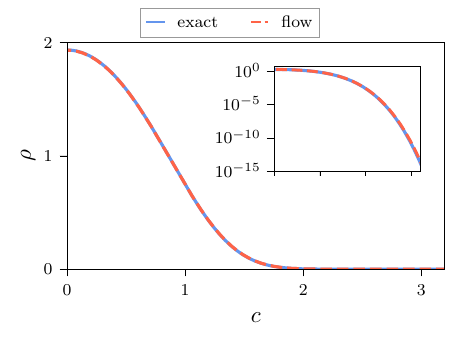}
    \caption{Comparison of the DoS computed with flow-based sampling to the exact solution for the zero-dimensional model.}
    \label{fig:site-rho}
\end{figure}

\begin{figure*}
	\centering
	\subfloat[]{%
		\includegraphics{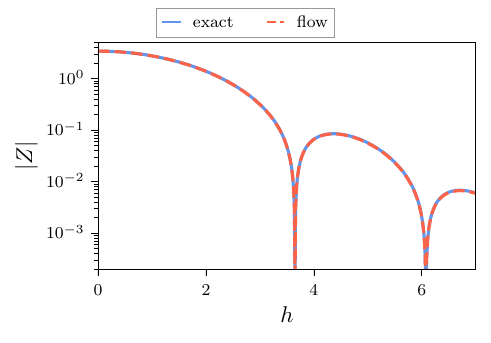}\label{fig:site-Z}}
	\hfill
	\subfloat[]{%
		\includegraphics{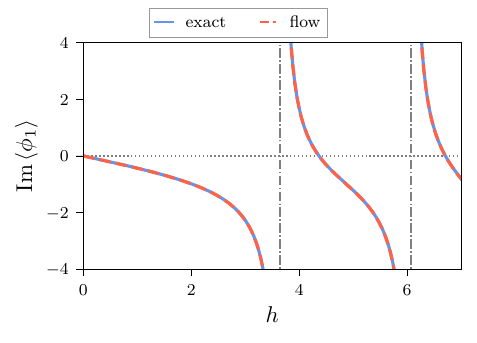}\label{fig:site-M}}%
	\caption{Comparison of the flow results to the exact solution for the zero-dimensional model: partition function (a) and average imaginary part of $\phi_1$ (b) as functions of $h$. The locations of the Lee-Yang zeroes in (a) are associated with the discontinuities of the observable in (b).}
\end{figure*}

For a first demonstration of our approach, we consider a zero-dimensional model of a single two-component scalar field $\phi = (\phi_1, \phi_2)$ with quartic self-interaction in an imaginary external field. The simplicity of this model facilitates a comparison to exact results. The action is
\begin{equation}
    S(\phi) = \frac{m^2}{2} \left( \phi_1^2 + \phi_2^2 \right) + \frac{\lambda}{4} \left( \phi_1^2 + \phi_2^2 \right)^2 + \im h \phi_1\ ,
\end{equation}
where $\phi_1, \phi_2, m^2, \lambda, h \in \mathbb{R}$. Accordingly, we can identify $X(\phi) \equiv \phi_1$ and
\begin{equation}
\begin{aligned}
    S_{c,P}(\phi) \equiv \frac{m^2}{2} \left( \phi_1^2 + \phi_2^2 \right) + \frac{\lambda}{4} \left( \phi_1^2 + \phi_2^2 \right)^2 \\
    + \frac{P}{2} \left(c - \phi_1 \right)^2 + \log \mathcal{N} \ .
\end{aligned}
\end{equation}

We train a normalizing flow using this target action with $m^2 = 1, \lambda = 1, P = 1000$. The context functions $s,t$ are implemented as fully-connected networks where the condition $c$ is provided as an additional input dimension; see \Cref{app:site-details} for further model and training details. The offset in this case is simply $\bar{\phi}(c) = (c, 0)$ such that $X(\bar{\phi}(c)) \equiv \bar{\phi}_1(c) = c$, as required by the construction of \Cref{sec:flow-dos}.

For the purpose of this proof-of-principle study, we simply assume for the remainder of this work that $\rho_P \approx \rho$ for sufficiently large $P$ and use \Cref{eq:rho-Z,eq:rho-obs} instead of \Cref{eq:rho-P-Z,eq:rho-P-obs}. While the accuracy in reproducing the exact results in this case completely justifies this assumption, we emphasize that this is an approximation and one should generally extrapolate $P \longrightarrow \infty$ more carefully. The particular value of $P$ used here has been adopted from~\cite{Borsanyi:2021gqg} for simplicity, and is observed to lead to reasonable results in the present setting. However, in general, it should be determined from a careful analysis of the trade-off between the accuracy of the DoS and the required computational effort, as larger values can lead to increasing autocorrelations in the case of HMC by forcing a reduction of the step size, as well as a degradation of the estimator variance in the case of our flow-based approach. 

The DoS computed with flow-based sampling is compared against the exact result in \Cref{fig:site-rho}, conclusively demonstrating the correctness of our approach across several orders of magnitude. Furthermore, the partition function $Z(h)$ is accurately reproduced, as shown in \Cref{fig:site-Z}. In particular, the locations of the first two Lee-Yang zeroes can be clearly identified. They are associated with discontinuities in the average imaginary part of $\phi_1$, which is also accurately determined with our method as shown in \Cref{fig:site-M}.

\subsection{One- and Two-Dimensional Models}

In order to verify that our approach also works beyond the rather trivial zero-dimensional setting, we consider actual low-dimensional lattice models of the two-component scalar field theory described above. Working in lattice units, the associated action in $d$ dimensions is defined as
\begin{equation}
\begin{aligned}
    S(\phi) &= \sum_{n \in \Lambda}  \Bigg(\frac{1}{2} \sum_{\mu = 1}^d \left|\phi(n) - \phi(n + \hat{\mu})\right|^2 \\
    &+ \frac{m^2}{2} \left| \phi(n) \right|^2 + \frac{\lambda}{4} |\phi(n)|^4 + \im h \phi_1(n) \Bigg) \ ,
\end{aligned}
\end{equation}
where $\phi(n) = \big( \phi_1(n), \phi_2(n) \big)$, $\Lambda$ is the set of all lattice sites given by integer-valued coordinates $n$, $\hat{\mu}$ denotes the unit vector in direction $\mu$, and we assume periodic boundary conditions in all directions. The action for each individual site is essentially equivalent to the zero-dimensional model, differing only in the additional kinetic term. Accordingly, we can identify
\begin{equation}
    X(\phi) \equiv \sum_{n \in \Lambda} \phi_1(n)
\end{equation}
as well as
\begin{equation}
\begin{aligned}
    S_{c,P}(\phi) &\equiv \sum_{n \in \Lambda}  \Bigg(\frac{1}{2} \sum_{\mu = 1}^d \left|\phi(n) - \phi(n + \hat{\mu})\right|^2 \\
    &+ \frac{m^2}{2} \left| \phi(n) \right|^2 + \frac{\lambda}{4} |\phi(n)|^4 \Bigg) \\
    &+ \frac{P}{2} \Big( c - \sum_{n \in \Lambda} \phi_1(n) \Big)^2 + \log \mathcal{N} \ .
\end{aligned}
\end{equation}

We train flows using this target action with $m^2 = 1, \lambda = 1, P = 1000$ for one- and two-dimensional lattices of size 8 and ${4 \times 4}$, respectively. To enforce equivariance under translations, the context functions $s,t$ are implemented as convolutional neural networks with the condition $c$ provided as an additional input channel. The offset in this case is $\bar{\phi}(c) = (c / |\Lambda|, 0)$, where $|\Lambda|$ denotes the total number of lattice sites, such that $X(\bar{\phi}(c)) = c$. In order to provide conventional baseline results, we employ HMC with the same target action and value for $P$, subsequently reconstructing $\rho_P$ up to an overall factor using the numerical integration method described in \Cref{sec:flow-dos}; see \Cref{app:lattice-details} for further details on model, training, and simulation.

\begin{figure*}
	\centering
	\subfloat[]{%
		\includegraphics{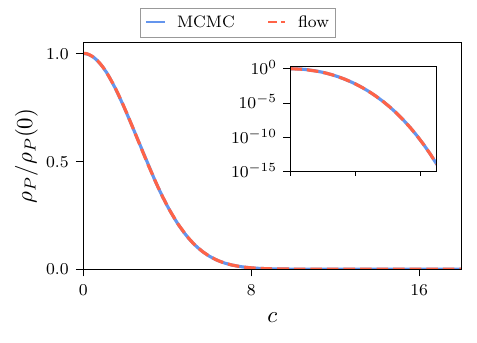}}
	\hfill
	\subfloat[]{%
		\includegraphics{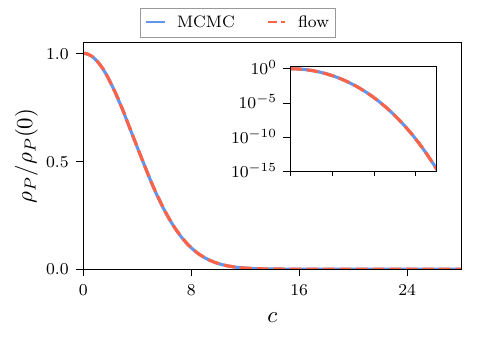}}%
	\caption{Comparison of the flow results for the normalized DoS to the reconstructions from MCMC calculations for the one- (a) and two-dimensional (b) models with $P = 1000$.}
    \label{fig:lattice-rho}
\end{figure*}

\Cref{fig:lattice-rho} compares $\rho_P(c) / \rho_P(0)$ (i.e.\ normalized to 1 at $c = 0$) obtained with flow-based sampling to the MCMC baseline. Similar to the zero-dimensional case, the results accurately reproduce the conventional computation, thereby confirming that our approach also works here as intended. We emphasize again that for the MCMC baseline, the reconstruction of the DoS at some point $c \neq 0$ by numerical integration requires precise knowledge of $\partial_c \log \rho_P(c)$ from $0$ to $c$. In contrast, with the flow-based approach, the DoS can be independently probed at arbitrary points because it is computed directly.

\section{Summary and Outlook}
\label{sec:summary}

In this paper, we apply flow-based sampling to the DoS approach to complex action problems. Specifically, we show that normalizing flows can be used to compute the DoS directly, thereby disposing of the need to reconstruct it from measurements of a derivative quantity through MCMC calculations. We demonstrate this method in the context of simple models with imaginary external fields, confirming the correctness and accuracy of our approach.

Due to the conceptual and practical differences between the flow-based and conventional strategies, an in-depth comparison of the computational cost is not straightforward and beyond the scope of this proof-of-principle study. We note that reaching the same level of accuracy in the final result using the flow-based approach is significantly cheaper in practice on the same hardware for the particular problem setting investigated in this work. This is likely due to the embarrassingly parallel sampling instead of the sequential evaluation of Markov chains. Furthermore, because of the numerical integration, the conventional ansatz may generally require higher precision in order to achieve an accurate reconstruction, whereas with the flow-based method the DoS can be directly probed at arbitrary points. However, it is unclear a priori how the upfront cost of training the flow compares against thermalizing the Markov chains. In general, the overall scaling of the computational cost required to achieve constant performance at different lattice volumes and action parameters depends strongly on the particular target and flow architecture, as well as the anticipated number of generated field configurations. Disentangling the various contributions to the total cost and defining thresholds for performance gains is a highly non-trivial task; see also~\cite{Abbott:2022zsh} for a detailed discussion of this issue. For DoS calculations as considered in the present work, the strong localization of the target density with increasing values of the parameter $P$ is expected to further increase the difficulty of the modeling task, thereby compounding the typically already poor volume scaling. Nevertheless---independently of how the cost actually scales in practice---depending on the particular scientific goals of such a calculation, the intrinsic advantages of the flow-based approach described in this work may well justify any additional expenses, and motivate further exploration in this direction.

In the future, we are interested in extending the present approach to higher dimensions, larger volumes, and fields with more components. This may be informative for the study of an approximate model of QCD near the second order phase transition where the external field plays the role of the quark mass~\cite{Attanasio:2021tio}. In particular, computing the DoS could help to constrain the location of the Lee-Yang edge singularity. In this context, it may also be worthwhile to implement equivariance of the flow under the residual $O(N-1)$ symmetry in order to better match the symmetries of the target distribution. Further interesting avenues include the relativistic Bose gas at finite chemical potential~\cite{Aarts:2008wh, Bongiovanni:2016jdj, Francesconi:2019nph} as well as the application to gauge theories via gauge-equivariant flows~\cite{Kanwar:2020xzo, Boyda:2020hsi}, such as e.g.\ $U(1)$ gauge theory with a topological term~\cite{Gattringer:2015eey} or QCD in the heavy-dense limit~\cite{Garron:2016noc}.

Beyond the aim to resolve complex action problems, the DoS can of course also be employed to compute observables for theories with purely real actions. While (unless one is interested in certain thermodynamic quantities) there is typically no reason to replace a standard MCMC algorithm with a generically much more expensive method, such an approach may be particularly useful for the treatment of ergodicity problems~\cite{Langfeld:2022uda}, since the target distribution can be mapped out explicitly in regions of configuration space that are otherwise pathologically under-sampled. Hence, the approach presented in this work also constitutes a promising ansatz for circumventing issues such as topological freezing and critical slowing down via flow-based methods, complementary to the more commonly investigated independence as well as hybrid sampling strategies.


\begin{acknowledgements}
This work is funded by the Deutsche Forschungsgemeinschaft (DFG, German Research Foundation) under Germany's Excellence Strategy EXC 2181/1 - 390900948 (the Heidelberg STRUCTURES Excellence Cluster) and the Collaborative Research Centre SFB 1225 (ISOQUANT). JMU is supported in part by the U.S.\ Department of Energy, Office of Science, Office of Nuclear Physics, under grant Contract Number DE-SC0011090. This work is funded by the U.S.\ National Science Foundation under Cooperative Agreement PHY-2019786 (The NSF AI Institute for Artificial Intelligence and Fundamental Interactions, \url{http://iaifi.org/}). The code used for the results in this paper is based on~\cite{Albergo:2021vyo}.
\end{acknowledgements}


\appendix

\section{Implementation details}
\label{app:details}

\subsection{Zero-dimensional model}
\label{app:site-details}

The flow for the zero-dimensional model consists of 16 affine coupling layers with context functions being fully-connected networks featuring three hidden layers with 64 neurons each. As activation functions we choose LeakyReLU~\cite{xu2015empirical} between layers together with a Tanh activation after the final layer. Each network has two input neurons, one for the frozen variable (either $\phi_1$ or $\phi_2$ depending on the layer) and the other one for the condition $c$; as well as two output neurons providing the values for $s,t$ in \Cref{eq:affine}. For the training, we apply the Adam optimizer with a learning rate of 1e--3 and a batch size of 10k, with a total of 5k gradient updates. In order to compute $\rho(c)$, 10k samples are drawn for each value of $c$, with a spacing of $\varDelta c = 0.01$.\\[1ex]

\subsection{One- and two-dimensional models}
\label{app:lattice-details}

For the one- and two-dimensional models, the flow also consists of 16 affine coupling layers. Here, the context functions are convolutional neural networks featuring two hidden layers with eight channels each, using intermediate LeakyReLU activations and a final Tanh activation as in the zero-dimensional case. Each network has three input channels, two for the frozen subsets of $\phi_1,\phi_2$ (determined by alternating checkerboard masking, see e.g.~\cite{Albergo:2021vyo}) and one for the condition $c$; as well as two output channels providing the values for $s,t$ in \Cref{eq:affine}. The conditional input is constructed with the same dimensions as a single component of $\phi$ with $c$ evenly distributed across all sites, i.e.\ with values of $c / |\Lambda|$ on each site where $|\Lambda|$ is the total number of lattice points. For the training, we apply the Adam optimizer with a learning rate of 1e--3 and a batch size of 1k, with a total of 50k gradient updates. In order to compute $\rho_P(c)$, 1e7 samples are drawn for each value of $c$, again with a spacing of $\varDelta c = 0.01$.

For the conventional MCMC calculations, we use HMC with a step size of 0.02 and 50 steps per trajectory for the one-dimensional as well as a step size of 0.01 and 100 steps for the two-dimensional case. This results in acceptance rates of roughly 60--90\%, with the highest values generally observed around $c = 0$ and decreasing rates for larger $c$. For each $c$, we run 10k Markov chains in parallel, where in each chain the first 1k steps are discarded for equilibration. Subsequently, the chains are evaluated for 100k steps and every 10th configuration is recorded, resulting in a total of 1e8 configurations for each value of $c$, using a spacing of $\varDelta c = 0.01$ as before. As described in the main text, $\rho_P(c) / \rho_P(0)$ is reconstructed from $\partial_c \log (\rho_P(c) / \rho_P(0))$ using the trapezoidal rule and exponentiating the resulting values for $\log (\rho_P(c) / \rho_P(0))$.

\bibliographystyle{utphys}
\bibliography{main}

\end{document}